\documentclass{article}
\usepackage[T1]{fontenc} 
\usepackage[utf8]{inputenc} 
\usepackage{ismir,amsmath,cite,url}
\usepackage{graphicx}
\usepackage{color}

\newcommand{\algoname}[1]{\textsc{{#1}}}
\newcommand{\ours}{\algoname{IPSim}}
\newcommand{\ldsd}{\algoname{LDSD}}
\newcommand{\random}{\algoname{Rnd}}
\newcommand{\countsegues}{\algoname{Count}}
\newcommand{\rarity}{$\mathit{rarity}$}
\newcommand{\unpop}{$\mathit{unpopularity}$}
\newcommand{\shortness}{$\mathit{shortness}$}
\newcommand{\interestingness}{$\mathit{interestingness}$}
\usepackage{amssymb}
\usepackage[normalem]{ulem}
\usepackage{tikz}
\usetikzlibrary{arrows.meta}

\title{An interpretable music similarity measure based on path interestingness}

\twoauthors
 {Giovanni Gabbolini} {Insight Centre for Data Analytics \\
School of Computer Science \& IT \\
University College Cork, Ireland \\
{\tt giovanni.gabbolini@insight-centre.org}}
 {Derek Bridge} {Insight Centre for Data Analytics \\
School of Computer Science \& IT \\
University College Cork, Ireland \\
{\tt d.bridge@cs.ucc.ie}}

\def\authorname{G. Gabbolini and D. Bridge}

\begin{document}

\maketitle
\begin{abstract}
We introduce a novel and interpretable path-based music similarity measure. Our similarity measure assumes that items, such as songs and artists, and information about those items are represented in a knowledge graph. We find paths in the graph between a seed and a target item; we score those paths based on their interestingness; and we aggregate those scores to determine the similarity between the seed and the target. A distinguishing feature of our similarity measure is its interpretability. In particular, we can translate the most interesting paths into natural language, so that the causes of the similarity judgements can be readily understood by humans. We compare the accuracy of our similarity measure with other competitive path-based similarity baselines in two experimental settings and with four datasets. 
The results highlight the validity of our approach to music similarity, and demonstrate that path interestingness scores can be the basis of an accurate and interpretable similarity measure.
\end{abstract}

\section{Introduction}\label{sec:introduction}

\setlength{\tabcolsep}{5pt}
\begin{table}[t] \begin{tabular}{ | l|l|} \hline Path (in natural language) & Int. \\ \hline
 George Jones was married to Tammy Wynette. & 0.60 \\ \hline
 \begin{tabular}{@{}lp{5cm}l}
     Tammy Wynette was the parent of the \\
     artist Georgette Jones and George Jones \\
     was also the parent of the same artist.
\end{tabular} & 0.56 \\ \hline
\begin{tabular}{@{}lp{5cm}l}
     Tammy Wynette wrote ``Our Private Life''  \\
     and George Jones also wrote the same song. 
\end{tabular}
   & 0.45 \\ \hline
\begin{tabular}{@{}lp{5cm}l}
    Tammy Wynette has made country music,  \\
     and so did George Jones.
\end{tabular}  & 0.34 \\ \hline
\begin{tabular}{@{}lp{5cm}l}
    Tammy Wynette was based in United States and  \\
     George Jones was based in the same country. 
\end{tabular}
   & 0.31 \\ \hline
   \begin{tabular}{@{}lp{5cm}l}
    Tammy Wynette was a solo artist, \\
     and George Jones was a solo artist. 
\end{tabular}
     & 0.29 \\ \hline
   \end{tabular} \caption{
   An example of the working of our proposed similarity measure. In this example, it is given the artists Tammy Wynette and George Jones as seed and target items.
}
   \label{table:case_study}
   \end{table}

The concept of similarity is central to music information retrieval, as it underpins important applications like recommender systems and visualisation in user interfaces \cite{Knees2016}. In recent years, there has been a surge in the use of knowledge graphs to solve various information retrieval tasks. For example, knowledge graphs have been applied to recommender systems \cite{Guo2020} and question answering \cite{Bordes2014}.
In this work, we use knowledge graphs to gauge music similarity. We introduce a novel path-based similarity measure. We represent items, such as songs and their artists, and information about those items in a knowledge graph; we find paths in the graph between a seed and a target item; we score those paths; and we aggregate the scores to determine the similarity between the seed and the target.

We propose to score paths based on interestingness. Interestingness is introduced in \cite{Gabbolini2021} as a way of distinguishing more interesting from less interesting paths in a knowledge graph. In \cite{Gabbolini2021}, the authors use what they call \textit{segues} to connect a song to the next song, e.g.\ in a playlist. Segues are translations into natural language of interesting paths in a knowledge graph. 
We also employ the path-to-text module from \cite{Gabbolini2021}, useful to translate paths to natural language. In this paper, we are extending \cite{Gabbolini2021} by showing that these paths and their interestingness scores can be the basis of an accurate and interpretable similarity measure.


For example, given the artists Tammy Wynette and George Jones as seed and target, our algorithm finds six paths between the two items, then it assigns an interestingness score to the paths, and it aggregates the scores to determine the similarity. It can convert the paths to natural language for display to a user. We report in Table \ref{table:case_study} its natural language translations of the paths (Path column) and the interestingness scores (Int.\ column).

A distinguishing feature of our similarity measure is its interpretability. Miller defines interpretability as the degree to which a human can understand the cause of a decision \cite{Miller2019}. According to this definition, our similarity measure is interpretable. Users can understand the causes of a similarity score by looking at natural language translations of the paths used to compute it, ranked by their interestingness.
For example, a user can understand why our algorithm thinks that Tammy Wynette and George Jones are similar, when presented with the ranked list of explanations in Table \ref{table:case_study}. Notice how the interestingness score intuitively reflects the interestingness of the paths. For example, two artists being married has higher score than two artists both being solo singers.

Music similarity measures are usually not interpretable, e.g.\ see  \cite{Knees2016book}. This may be related to a system-centric research protocol, which is common in the field, 
and consists of laboratory experiments conducted without end-user involvement, e.g.\ the evaluation of algorithms on digital databases \cite{Schedl2013}. This protocol is pragmatic but it may have contributed to widening the ``semantic gap'' in music similarity measures, which sets a bound on user satisfaction \cite{Knees2016}. 
The term ``semantic gap'' refers to the discrepancy between high-level perception of similarity by humans and the low-level numerical data used by algorithms \cite{Ponceleon2011}. 
For example, a user might perceive two songs as similar because of their lyrics, while an algorithm might be limited to using abstract statistical descriptors operating on the audio signals. 
One way of alleviating the semantic gap is the introduction of mid-level features. Mid-level features either combine low-level features or extend them to ones that incorporate additional, higher-level knowledge \cite{Knees2016}. Another way of reducing the gap is the adoption of interpretable similarity measures. An interpretable similarity measure can narrow the gap between the low perceptual level of an algorithm, which makes a decision about similarity, and the high perceptual level of a human, by providing an explanation for the human to understand the decision. 


The remainder of the paper is organised as follows: in Section \ref{sec:related_works}, we frame our similarity measure in the literature of the subject; in Section \ref{sec:method}, we formalise our similarity measure; and, in Section \ref{sec:experiments}, we validate our similarity measure in four experimental conditions. The source code supporting this study is freely available.\footnote{https://github.com/GiovanniGabbolini/ipsim} 

\section{Related work}\label{sec:related_works}

The concept of similarity is central to Music Information Retrieval.
Over the years, researchers have proposed many ways of measuring music similarity, which can be categorised based on the data they use. We might, for example, categorise them into two: content-based approaches that use the audio signal; and context-based approaches that use information about artists and the pieces of music themselves, e.g.\ their lyrics, listening logs, and so on \cite{Knees2013}. We refer the reader to the book by Knees \& Schedl \cite{Knees2016book} for an extensive review of content-based and context-based approaches.
The similarity measure that we propose in this paper is context-based.

Some of the context-based approaches make use of knowledge graphs.
These approaches represent entities from the music domain and information about those entities as nodes in a graph, and they use graph structure to gauge similarity. (We define knowledge graphs more formally in Section \ref{sec:kg}.)

Many knowledge graph-based approaches to the measurement of similarity are path-based. They use the existence of paths between the seed and the target entities to determine similarity.
Path-based algorithms usually score paths based on some criteria, and then aggregate path scores to determine similarity.
For example, in \cite{Leal2012} Leal et al.\ describe Shakti. Shakti scores each path by summing hand-crafted node weights and edge weights, and then dividing by the cube of the path length. Shakti only considers paths whose length falls below a fixed threshold. In \cite{Strobin2013}, Strobin \& Niewiadomski extend Shakti by setting node and edge weights using a genetic algorithm. This extension evolves different sets of weights indexed by path length, and considers paths of any length. They adjust node weights by considering centrality, i.e.\ the number of in-going and out-going edges of a node.

Passant proposes another path-based algorithm, called LDSD \cite{Passant2010}. LDSD considers only paths of length up to two. It assigns low scores to paths whose type is common in the context of the seed and target entities. Piao \& Breslin extend LDSD in \cite{Piao2016}. They propose four variants of LDSD, which correct the path scoring mechanism with global information, such as the frequency of path types in the whole knowledge graph.
In a comparative study, Piao et al.\ show that LDSD largely outperforms Shakti in accuracy \cite{Piao2016resim}.

The similarity measure that we propose in this paper is also path-based. It is different from the works in the literature as we employ a novel scoring mechanism for paths, based on their interestingness. 

As well as path-based approaches to the measurement of similarity, there are also embedding-based approaches. These work by representing the knowledge graph in a dense, low-dimensional feature space, which tries to preserve as much of the graph's structural information as possible. There are several ways to learn knowledge-graph embeddings. We refer the reader to \cite{Dai2020} for a survey. Once the embeddings are learned, the similarity between entities can be computed on their vector representations using, for example, cosine similarity.

Path-based and embedding-based approaches to the measurement of similarity differ in their interpretability. In \cite{Du2019}, Du et al.\ provide a characterization of approaches based on their potential for interpretability. They introduce the concept of \textit{intrinsic interpretability}, which is achieved by constructing self-explanatory models. 
Path-based approaches typically exhibit intrinsic interpretability, since the paths used to gauge similarity can be considered to be self-explanatory. Embedding-based approaches do not exhibit intrinsic interpretability, since the embeddings used to gauge similarity cannot be considered to be self-explanatory. As a consequence, path-based approaches are interpretable after the translation of the paths to natural language, while embedding-based approaches can become interpretable only after the introduction of a post-hoc model, built to generate explanations \cite{Du2019}. 

Path-based approaches are not the only category of similarity measures to feature intrinsic intepretability.
For example, if items are described by sets of tags, then Jaccard similarity over tags also exhibits intrinsic intepretability. 
The similarity can be explained by showing the tags shared between the seed and the target. Even so, we find a scarcity of interpretable similarity measures in the literature; the book \cite{Knees2016book} contains very few, for example. 


We mention that Passant's LDSD is also interpretable, since Passant implements a mechanism to translate paths to natural language \cite{Passant2010} .



\section{Method} \label{sec:method}


As we have already indicated, our approach to the measurement of similarity of entities in knowledge graphs is path-based. It builds on work described in \cite{Gabbolini2021}, where the authors introduce interestingness as a way of distinguishing more interesting from less interesting paths in a knowledge graph.

In the following, we first introduce some preliminary concepts, then we review \cite{Gabbolini2021}'s definition of interestingness, and finally we introduce our similarity measure.

\subsection{Preliminaries} \label{sec:kg}

Both our work and \cite{Gabbolini2021} make use of a knowledge graph $G$ as an abstract representation for items and information about those items. In the experiments that we report later in this paper, items are artists; in \cite{Gabbolini2021}, items are songs. But the approach is domain-independent: items can be any entities of interest.

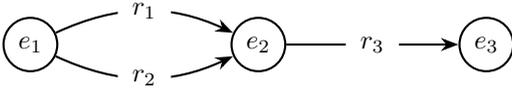
\begin{figure}
\begin{tikzpicture}
\begin{scope}[every node/.style={circle,thick,draw}]
    \node (A) at (0,0) {$e_1$};
    \node (B) at (3,0) {$e_2$};
    \node (C) at (6,0) {$e_3$};
\end{scope}

\begin{scope}[>={Stealth[black]},
              every node/.style={fill=white,circle},
              every edge/.style={draw=black, thick}]
    \path [->] (A) edge[bend left=25]  node {$r_1$} (B);
    \path [->] (A) edge[bend right=25] node {$r_2$} (B);
    \path [->] (B) edge node {$r_3$} (C);
\end{scope}
\end{tikzpicture}
\caption{A simple knowledge graph.} 
\label{fig:kg}
\end{figure}

\begin{description}
\item[\textnormal{A} knowledge graph] is a set of triples $G=\{(e,r,e')\,|\,e,e' \in E, r \in R\}$, where $E$ and $R$ denote, respectively, the sets of entities (nodes) and relationships (edges). 
For example, the knowledge graph of Figure \ref{fig:kg} contains three triples, $\{(e_1, r_1, e_2), (e_1, r_2, e_2), (e_2, r_3, e_3)\}$. A special subset of entities $I \subseteq E$, are the items. Every entity has a \textit{type} and a \textit{value}.
For example, if an entity represents an artist, then its type is ``artist'' and its value is the URI of that artist.
Every edge (relationship) has a \textit{type} also. For example, an edge that connects two artists who have collaborated will have ``collaborated with'' as its type.
\item[\textnormal{A} path] $p$ in $G$ is an ordered list of entities and edges in $G$, $p=[e_1, r_1, ..., r_{n-1}, e_n]$ where each triple in $p$ must be in $G$. For example, with reference to the knowledge graph of Figure \ref{fig:kg}: $[ e_1, r_1, e_2, r_3, e_3 ]$ is a path; $[ e_1, r_1, e_2, r_1, e_3 ]$ is not a path, as the triple $(e_2, r_1, e_3)$ is not in the knowledge graph. The $\mathit{type}$ of $p$ is the ordered concatenation of the entity and edge types in $p$. 
\end{description}
The function $\mathit{paths(i_1, i_2)}$ finds the paths in $G$ from the item $i_1 \in I$ to the item $i_2 \in I$, without visiting any other item in $I$ and without cycles. In other words, if items are songs, for example, then the function finds paths between pairs of songs, where the paths are not allowed to contain any intermediate songs.

\subsection{Interestingness} \label{subsubsect:interestingness}



In \cite{Gabbolini2021}, the interestingness of a path in a knowledge graph is defined as a weighted combination of three heuristics:

\begin{description}
 \item[Rarity] 
 Let $T$ be the set of all path types in $G$; and let $f(t)$ be the number of paths in $G$ that are  of type $t$. The \rarity{} of a path $p$ is defined as:
\[
     \mathit{rarity}(p)=1-\frac{f(\mathit{type}(p))}{\max_{t \in T} f(t)}
 \]

\item[Unpopularity]
Let $\mathit{edgeset}(e)$ be the set of in-going and out-going edges to and from an entity $e \in E$ in $G$.
The centrality of an entity $e$ is:

\[
\begin{split}
    \mathit{centrality}(e)= \min \left( 1, \frac{|\mathit{edgeset}(e)|}{\underset{e' \in E}{\mathit{median}} |\mathit{edgeset}(e')|} \right) , \\
    \text{given} \,\, \mathit{type}(e')=\mathit{type}(e)
\end{split}
\]

The \unpop{} of a path $p$ is defined as:
\[
    \mathit{unpopularity}(p)=1-\min_{e \in p \cap E}(\mathit{centrality}(e))
\]

\item[Shortness] 
If we define $\mathit{length}(p)$ as the number of edges in path $p$,
then the \shortness{} of a path is given by:
\[
    \mathit{shortness}(p)=\frac{1}{\mathit{length}(p)}
\]

\end{description}
The $\mathit{interestingness}$ of a path $p$ in $G$ is given by:
\[
\begin{split}
    \mathit{interestingness}(p)=w_1 \mathit{rarity}(p) + \\
    w_2 \mathit{unpopularity}(p) + \\
    w_3 \mathit{shortness}(p)
\end{split}
\]
Its values range from zero to one.
$w_1, w_2, w_3$ are parameters to be tuned, subject to $w_1+w_2+w_3=1.$

In \cite{Gabbolini2021}, the authors run a user trial where participants are asked to evaluate segues. They find that \interestingness{} positively correlates with human perceptions of quality: \interestingness{} is high for segues perceived as high-quality, and vice versa. This lends credence to the idea of using their definition of interestingness in the similarity measure that we are proposing in this paper.

\subsection{Similarity} \label{subsect:algorithm}



We define the similarity between two items $i_1 \in I$ and $i_2 \in I$ as an aggregate of the interestingness of the paths between them:
\[
sim(i_1, i_2)= \sum_{p \in paths(i_1, i_2)} \mathit{interestingness}(p) \, \mathcal{I}(p, n)
\]
where $\mathcal{I}$ is an indicator function: $\mathcal{I}(p,n)=1$ if $\mathit{length}(p) \leq n$, and $0$ otherwise. $n$ is a parameter whose value determines whether the aggregation restricts attention to shorter paths or whether it considers all paths ($n = \infty$). It is not clear \textit{a priori} whether it is beneficial to put a limit on path length and, if so, what limit to use. For example, \cite{Passant2010} use only paths of length up to three, while \cite{Strobin2013} uses all paths. In our case, we leave the choice open by introducing the parameter $n$ to be tuned. The interestingness weights $w_1$, $w_2$, $w_3$ are also parameters to be tuned.

In the following, we refer to our similarity measure as \ours{} (\textbf{I}nteresting \textbf{P}ath \textbf{Sim}ilarity).
\ours{} is domain-independent. In the rest of this paper, we focus on the music domain, and in particular the case where items are music artists. The reason we focus on artist similarity is that it is a common choice in the literature and there is an amount of data to use in offline experiments. 

\section{Experiments}\label{sec:experiments}

We provide an evaluation for \ours{} in the case of music artists. 
In this evaluation, we are particularly interested in evaluating the accuracy that \ours{} is able to achieve.

\subsection{Knowledge graph}

We represent an artist with their MusicBrainz URI.\footnote{\url{https://musicbrainz.org/}} This representation can be easily changed with only minor modifications to the rest of our implementation, e.g.\ if artists were instead represented by their Spotify URIs.

Our implementation uses a knowledge graph with 30 distinct node types and 205 distinct edge types.
It is built with data that we harvest from two  resources:

\begin{description}
 \item[MusicBrainz] We use MusicBrainz as the main source of factual data. We exploit the MusicBrainz APIs.\footnote{\url{https://python-musicbrainzngs.readthedocs.io/en/v0.7.1/}} They allow us to navigate the MusicBrainz database, and offer entity-linking functionalities. We mine different sorts of factual data, ranging from the birth places of the artists to the genres that they play. 

\item[Wikidata] We use Wikidata as an additional source of factual data.\footnote{\url{https://wikidata.org/}} There exists a mapping from MusicBrainz URIs to Wikidata URIs, making it easy to use both resources. From Wikidata, we mine biographical data about artists that is not available in MusicBrainz, e.g.\ the awards that an artist has won.
\end{description}
We provide a complete description of entities and relationships that build up the knowledge graph that we use in this paper in the additional materials.\footnote{\url{https://doi.org/10.5281/zenodo.5121460}} 

\subsection{Experiment design}\label{subsec:experiments_design}
As highlighted by Knees \& Schedl \cite{Knees2016}, there is no agreement upon the best method to evaluate the quality of measures of music similarity. Every method has its own advantages and disadvantages. In this paper, we assess music similarity measures using two common experimental settings. The first uses a similar-artists ground-truth, and the second uses user-artist interaction histories.

\subsubsection{Using a ground-truth} \label{subsubsect:experiment_ground_truth}
This evaluation procedure replicates the one in \cite{Oramas2015}.
The procedure uses datasets that are composed of tuples of the form \textit{(seed artist, similar-artists ground-truth)}. In other words, for each seed artist, it gives a list of artists that have been independently judged to be similar to the seed, these artists being regarded as the ground-truth. We employ two datasets:
\begin{description}
    \item[\textit{MIREX}:] This dataset comprises 188 seed artists. The ground truth is based on similarity judgements expressed by experts during the MIREX Audio Music Similarity and Retrieval Task. 
    \item[\textit{LastFM-g}:] This dataset comprises 2336 seed artists. The ground truth is based on similarity judgements gathered from the Last.fm APIs.\footnote{\url{last.fm}} 
\end{description}
See \cite{Oramas2015} for additional 
details on these datasets.

To evaluate a similarity measure on a given dataset, we do the following.
For every seed artist, we use the similarity measure to score the other artists and then to rank them by decreasing similarity with the seed. Then, we compare this ranking \textit{@N} with the ground-truth from the dataset, according to the standard accuracy metrics \textit{nDCG} and \textit{Precision}, or \textit{Pr} for short.

\subsubsection{Using interaction histories} \label{subsubsect:experiment_interaction_histories}


This protocol uses a dataset that records how users have interacted with artists.
We denote with $U$, $I$ the sets of users and artists resp.\ and with $|U|$, $|I|$ their cardinalities. Lower case letters $u$, $i$ refer to $u \in U$, $i \in I$. In this case, the dataset is organised as a matrix $R \in \mathbb{R}^{|U| \times |I|}$. Each cell $r_{ui} > 0$ of $R$ accounts for a user-artist interaction. If a user $u$ is not known to have interacted with an artist $i$, then $r_{ui}$ is zero.

We employ two datasets of this kind:
\begin{description}
    \item[\textit{LastFM-h}:] In this Last.fm dataset (which is different from the one described earlier), $r_{ui}$ is the listening count by user $u$ of artist $i$ \cite{Cantador2011}.\footnote{\url{https://grouplens.org/datasets/hetrec-2011/}} We filter out users who have listened to fewer than five artists, and artists listened to by fewer than five users. We end up with 1877 users and 2828 artists. We convert listening counts to ratings on a 1--5 scale following the procedure given in  \cite{Celma2008}. 
    For every user, we hold-out at random 40\% of her interactions, and we keep the other interactions as training data.
     We further split the held-out interactions into two equal-sized parts, to form validation and test data. Finally, we remove the interactions where $r_{ui} < 4$ from the validation and test data so that they only contain artists that the users like.
    \item[\textit{Facebook}:] This dataset, from the Second Linked Open Data-enabled Recommender Systems Challenge, contains items liked by users in three domains: movies, books and music.\footnote{\url{https://lists.w3.org/Archives/Public/public-vocabs/2015Feb/0046.html}} 
    We focus 
    on music artists, by keeping only items of types: \textit{music\_artist} and \textit{music\_band}. We filter out artists liked by fewer than five users, and those whose MusicBrainz URI cannot be resolved. We end up with 52069 users and 4435 artists. In this case, $r_{ui} = 1$ if the user $u$ likes the artist $i$. For every user, we hold-out 40\% of her interactions at random, and we keep the others as training data. We further split the held-out interactions into two equal-sized parts, to form validation and test data.
\end{description}

For these two datasets, the evaluation procedure uses the similarity measure as a core component within a recommender system. The design of the recommender system is inspired by \cite{Piao2016}, chosen because of its heavy reliance on similarity, which is what we are trying to evaluate. We define the user profile of a user $u$ as the artists she has rated (in the training set):
\[
P(u)= \{ j | r_{uj} \neq 0 \}.
\]
Let $\mathit{NN}(k, i, P(u))$ be the set of \textit{k} most similar items in $P(u)$ to an artist $i$.
The predicted relevance score for a user $u$ and an artist $i$, indicated as $\overline{r_{ui}}$, is computed as:
\[
 \overline{r_{ui}} = 
 \begin{cases}
    0              & \text{if } i \in P(u) \\
    \sum_{j \in \mathit{NN}(k, i, P(u))}{r_{uj} \ sim(i,j)} & \text{if } i \not\in P(u)
\end{cases}
\]

For every user, we compute the relevance score $\overline{r_{ui}}$ for every artist $i \in I$, and rank the artists by decreasing score.  Then, we compare this ranking \textit{@N} and the held-out test data (or validation data, as appropriate), according to the standard accuracy metrics \textit{nDCG} and \textit{Precision}, or \textit{Pr} for short.
We take higher recommendation accuracy to be an indication of a higher quality similarity measure. 

\subsection{Baseline similarity measures}\label{subsec:baselines}

We include in our experiments three baselines. The first is a random algorithm (\random{}), useful to set a lower bound on the performance. It rates the similarity between two items to be a random number between zero and one. The other two (\countsegues{} and \ldsd{}) are path-based algorithms. 
By comparing our path-based approach to  other path-based similarities, we reduce the number of confounders, so we can investigate whether our idea of using interestingness as a scoring mechanism for paths is a good one.
In fact, the path-based similarity measures that we use in this experiment all use the same knowledge graph and the same paths, and differ only in the way the paths are scored and how the scores are aggregated.
\countsegues{} is a simple path-based algorithm that works by counting the number of paths between items. It is equivalent to substituting 1 for the argument to the sum in the definition of $\mathit{sim}$ given earlier. \ldsd{} is 
an accurate path-based algorithm that proposes 
a carefully-designed weighting heuristic for paths \cite{Passant2010}.


\subsection{Parameter tuning}\label{subsec:param_tuning}

\ours{} has some parameters, as indicated in Section \ref{subsect:algorithm}. Additionally, the experiment using listening histories introduces another parameter, i.e.\ the number of neighbours $k$ used by the recommender system, as described in Section \ref{subsubsect:experiment_interaction_histories}. We set the parameters with a grid-search, optimizing \textit{nDCG@10} on the validation data. For the experiment using a ground-truth, we tune the parameters of \ours{}. For the experiment using listening histories, we tune the parameters of \ours{} and we tune $k$ for each of \random{}, \countsegues{}, \ldsd{} and \ours{}. 
We report the optimal parameters found for \ours{} in Table \ref{table:parameters}. The optimal values of the parameter $k$ for \random{}, \countsegues{} and \ldsd{} are, respectively, $5$, $10$ \& $40$ in \textit{LastFM-h} and $1$, $1$ \& $7$ in \textit{Facebook}.

The optimal parameter configurations change from dataset to dataset.
This might be due to the different nature of the datasets. \textit{MIREX} features human judgements of artist similarity; the algorithmic judgments to be found in \textit{LastFM-g} might not completely agree with these. Schedl et al.\ \cite{Schedl2013} suggest that human perception of similarity is influenced by user factors, such as user context (e.g.\ mood) and user properties (e.g.\ musical training). User factors might have directly influenced the people who annotated \textit{MIREX}, but not the algorithm used to construct \textit{LastFM-g}. 
The remaining two datasets, \textit{LastFM-h} and \textit{Facebook}, gather another kind of data, i.e.\ interaction histories, 
but they are gathered in different ways. In \textit{LastFM-h}, unless the user intervenes, the next recommended song is played automatically, whereas, in \textit{Facebook}, users are interacting with artist fan pages in what is usually a more deliberate, conscious and considered way. 

In three of the four datasets, at least one of the three interestingness heuristics of \ours{} receives a weight of zero. We may find an explanation for this in the median number of paths between artists: 19 in \textit{MIREX}, 13 in \textit{LastFM-g}, and seven in both \textit{LastFM-h} and \textit{Facebook}. \textit{MIREX} is the only dataset where all three weights are non-zero, and is also the dataset with the highest median number of paths between artists. In general, the use of a more complex model (with non-zero weights for each component) might be beneficial only when the dataset refers to artists in the knowledge graphs for whom there exists a rich variety of paths, e.g.\ when considering popular artists.

\setlength{\tabcolsep}{5pt}
\begin{table} 
\centering
\begin{tabular}{|l|l|l|l|l|l|}
\hline
{} &         $w_1$ & $w_2$ & $w_3$ & $n$ & $k$ \\ \hline

\textit{MIREX}      &                   0.3 &                 0.1 &                   0.6 &                 2 & NA \\ \hline
\textit{LastFM-g} &                   0.0 &                 0.1 &                   0.9 &                 2  & NA \\ \hline
\textit{LastFM-h}        &                   0.0 &                 0.0 &                   1.0 &                 2 & 40  \\ \hline
\textit{Facebook}        & 0.9 & 0.0 & 0.1 & $\infty$ & $\infty$  \\ \hline
\end{tabular}
\caption{Optimal parameters found for \ours{}.}
 \label{table:parameters} 
 \end{table}

\subsection{Results}

\setlength{\tabcolsep}{5pt}
\begin{table} 
\begin{tabular}{|l|l|l|l|l|}
\hline
{} &         \textit{nDCG@5} &      \textit{nDCG@10} &           \textit{Pr@5} &        \textit{Pr@10} \\ \hline

\random{}      &                   0.023 &                 0.022 &                   0.026 &                 0.023 \\ \hline
\countsegues{} &                   0.065 &                 0.057 &                   0.062 &                 0.052 \\ \hline
\ldsd{}        &                   0.102 &                 0.094 &                   0.096 &                 0.088 \\ \hline
\ours{}        &  \textbf{0.136}$^{***}$ &  \textbf{0.114}$^{*}$ &  \textbf{0.130}$^{***}$ &  \textbf{0.101}$^{*}$ \\ \hline

\end{tabular}
\caption{Accuracy of similarity measures on the \textit{MIREX} dataset, as measured by \textit{nDCG} and \textit{Precision} at cutoffs five and ten. $^*$: $p$<0.05; $^{**}$: $p$<0.01; $^{***}$: $p$<0.001.}
 \label{table:accuracy_ground_truth_mirex} 
 \end{table}

\setlength{\tabcolsep}{4.8pt}
\begin{table} 
\begin{tabular}{|l|l|l|l|l|}
\hline
{} &        \textit{nDCG@5} &        \textit{nDCG@10} &          \textit{Pr@5} &          \textit{Pr@10} \\ \hline

\random{}      &                  0.002 &                   0.002 &                  0.002 &                   0.002 \\ \hline
\countsegues{} &                  0.037 &                   0.030 &                  0.033 &                   0.025 \\ \hline
\ldsd{}        &                  0.103 &                   0.086 &                  0.095 &                   0.075 \\ \hline
\ours{}        &  \textbf{0.111}$^{**}$ &  \textbf{0.094}$^{***}$ &  \textbf{0.103}$^{**}$ &  \textbf{0.083}$^{***}$ \\ \hline

\end{tabular}
\caption{Accuracy of similarity measures on the \textit{LastFM-g} dataset, as measured by \textit{nDCG} and \textit{Precision} at cutoffs five and ten. $^*$: $p$<0.05; $^{**}$: $p$<0.01; $^{***}$: $p$<0.001.}
 \label{table:accuracy_ground_truth_lastfm_g}
 \end{table}

\begin{table} 
\begin{tabular}{|l|l|l|l|l|}
\hline
{} &         \textit{nDCG@5} &      \textit{nDCG@10} &           \textit{Pr@5} &         \textit{Pr@10} \\ \hline

\random{}      &                   0.001 &                  0.000 &                   0.001 &                0.000 \\ \hline
\countsegues{} &                   0.015 &                  0.012 &                   0.014 &                0.010 \\ \hline
\ldsd{}        &  \textbf{0.025}$^{***}$ &  \textbf{0.020}$^{**}$ &  \textbf{0.022}$^{***}$ &                0.016 \\ \hline
\ours{}        &                   0.016 &                  0.017 &                   0.017 &  \textbf{0.017}$^{}$ \\ \hline

\end{tabular}
\caption{Accuracy of similarity measures on the \textit{LastFM-h} dataset, as measured by \textit{nDCG} and \textit{Precision} at cutoffs five and ten. $^*$: $p$<0.05; $^{**}$: $p$<0.01; $^{***}$: $p$<0.001.}
 \label{table:accuracy_listening_histories_lastfm_h}
 \end{table}
 
\setlength{\tabcolsep}{4.5pt}
 \begin{table} 
\begin{tabular}{|l|l|l|l|l|}
\hline
{} &         \textit{nDCG@5} &        \textit{nDCG@10} &           \textit{Pr@5} &          \textit{Pr@10} \\ \hline

\random{}      &                   0.001 &                   0.001 &                   0.001 &                   0.001 \\ \hline
\countsegues{} &                   0.017 &                   0.014 &                   0.016 &                   0.012 \\ \hline
\ldsd{}        &                   0.021 &                   0.018 &                   0.019 &                   0.017 \\ \hline
\ours{}        &  \textbf{0.024}$^{***}$ &  \textbf{0.021}$^{***}$ &  \textbf{0.023}$^{***}$ &  \textbf{0.020}$^{***}$ \\ \hline
\end{tabular}
\caption{Accuracy of similarity measures on the \textit{Facebook} dataset, as measured by \textit{nDCG} and \textit{Precision} at cutoffs five and ten. $^*$: $p$<0.05; $^{**}$: $p$<0.01; $^{***}$: $p$<0.001.}
 \label{table:accuracy_listening_histories_facebook}
 \end{table}

\begin{figure*}
 \centerline{
 \includegraphics[width=\linewidth]{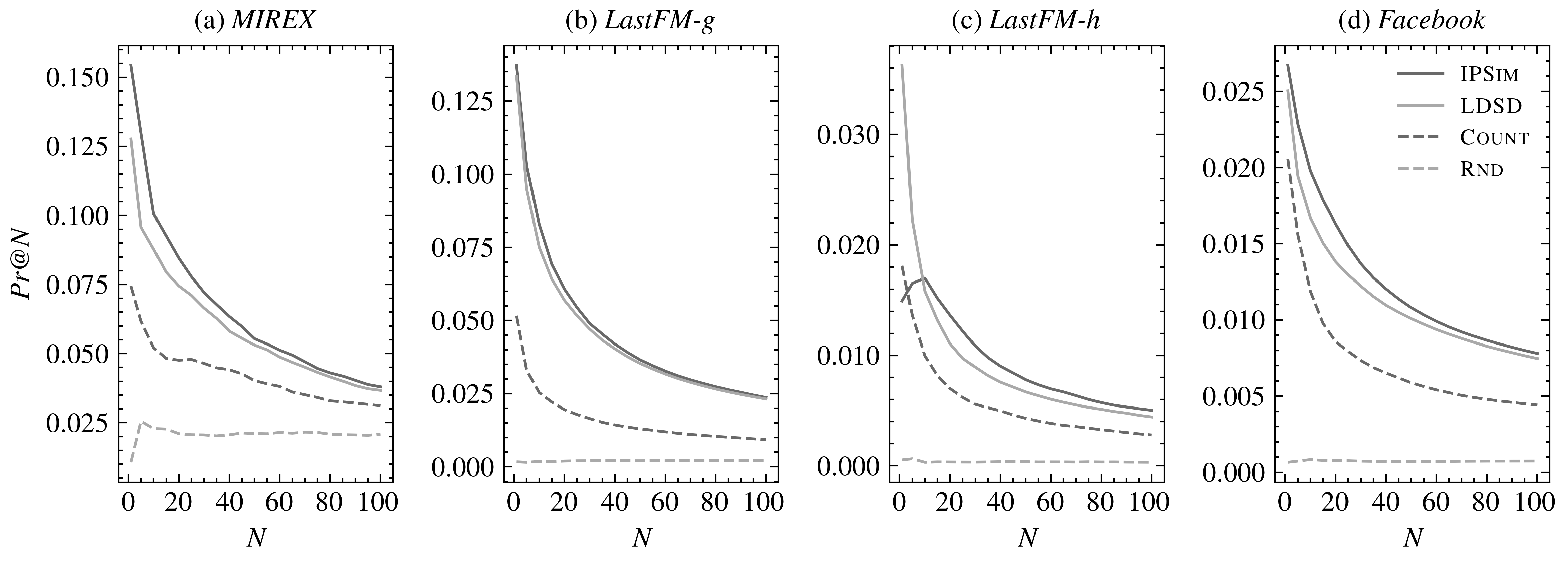}}
 \caption{Accuracy of similarity measures on the \textit{MIREX}, \textit{LastFM-g}, \textit{LastFM-h} and \textit{Facebook} datasets, as measured by \textit{Precision} at cutoffs from one to 100.}
 \label{fig:accuracy}
\end{figure*}

We conduct the experiments described in Section \ref{subsec:experiments_design}. We compare \ours{} against the baselines described in Section \ref{subsec:baselines} (\random{}, \countsegues{}, \ldsd{}). We set the parameters of the similarity measures and the recommender as described in Section \ref{subsec:param_tuning} using the validation data. We report the results on the test data. We verify the significance of differences between \ours{} and \ldsd{} with Wilcoxon signed-rank test.

Tables \ref{table:accuracy_ground_truth_mirex} \& \ref{table:accuracy_ground_truth_lastfm_g} report on the experiment that uses a ground-truth. \ours{} outperforms all the baselines in both the \textit{MIREX}
and \textit{LastFM-g} datasets. We notice that \ours{} scores accuracy at least double that of \countsegues{} in both datasets. The increase in performance with respect to \ldsd{} is statistically significant in both datasets. 

Tables \ref{table:accuracy_listening_histories_lastfm_h} \& \ref{table:accuracy_listening_histories_facebook} report on the experiment that uses listening histories. \ours{} is always more accurate than \countsegues{}. \ours{} always outperforms \ldsd{} in \textit{Facebook} but outperforms \ldsd{} in only one case in \textit{LastFM-g}, and the difference in this case is not statistically significant. 

We investigate more deeply how accuracy varies with the cutoff in Figure \ref{fig:accuracy}. For the four datasets, the figure shows the \textit{Precision} of the similarity measures as a function of the cutoff. The figure confirms that \ours{} has higher precision than \ldsd{} on \textit{LastFM-h} for cutoffs greater than ten, but the precision of \ours{} drops for cutoffs $< 10$. We also notice that, on \textit{MIREX}, \textit{LastFM-g} and \textit{Facebook}, \ours{} outperforms \ldsd{} for every cutoff ranging from one to 100. 

The results provide evidence that interestingness can be the basis of an accurate similarity measure.

\section{Conclusions and future work}
The results of the experiments highlight the validity of our approach to music similarity (at least in the case of music artists), and demonstrate that the interestingness scores of \cite{Gabbolini2021} can be the basis of an accurate and interpretable similarity measure.
Future work might include a comparison of the performance of our similarity measure against an even wider range of music similarity measures. We are also interested in using the proposed similarity measure in a related-item recommender system, that can meaningfully guide users in their exploration of the items by means of the interpretability of the similarity measure.

\section{Acknowledgments}
This publication has emanated from research conducted with
the financial support of Science Foundation Ireland under Grant
number 12/RC/2289-P2 which is co-funded under the European
Regional Development Fund. For the purpose of Open Access, the
authors have applied a CC BY public copyright licence to any Author
Accepted Manuscript version arising from this submission.

\bibliography{ISMIRtemplate}

\end{document}